\documentclass[twocolumn,showpacs,preprintnumbers,amsmath,amssymb,pra,superscriptaddress]{revtex4}
\usepackage{mathrsfs}
\usepackage{graphicx}
\usepackage{dcolumn}
\usepackage{bm}
\usepackage{amsmath}
\usepackage{amsfonts}
\usepackage{txfonts}

\begin{document}

\title{Coherent spin control by electromagnetic vacuum fluctuations}
\author{Jing Wang}
\affiliation{Department of Physics, The Chinese University of Hong Kong, Shatin, N.T., Hong Kong, China}
\affiliation{Department of Physics, Tsinghua University, Beijing 100084, China}
\author{Ren-Bao Liu} \thanks{Corresponding author. rbliu@cuhk.edu.hk}
\affiliation{Department of Physics, The Chinese University of Hong Kong, Shatin, N.T., Hong Kong, China}
\author{Bang-Fen Zhu}
\affiliation{Department of Physics, Tsinghua University, Beijing 100084, China}
\affiliation{The Institute of Advanced Study, Tsinghua University, Beijing 100084, China}
\author{L.~J.~Sham}
\affiliation{Department of Physics, University of California San Diego, La Jolla, California 92093-0319, USA}
\author{D.~G.~Steel}
\affiliation{The H. M. Randall Laboratory of Physics, University of Michigan, Ann Arbor, Michigan 48109, USA}
\date{\today}

\begin{abstract}
In coherent control, electromagnetic vacuum fluctuations usually cause coherence loss
through irreversible spontaneous emission. However, since the dissipation via emission
is essentially due to correlation of the fluctuations, when emission ends in a superposition of
multiple final states, correlation between different pathways
may build up if the ``which-way'' information is not fully resolved
(i.e., the emission spectrum is broader than the transition energy range).
Such correlation can be exploited for spin-flip control in a $\Lambda$-type
three-level system, which manifests itself as an all-optical spin echo in nonlinear optics
with two orders of optical fields saved as compared with stimulated Raman processes.
This finding represents a new class of optical nonlinearity induced by electromagnetic vacuum fluctuations.
\end{abstract}


\maketitle

\section{Introduction}
\label{introduction}
Electromagnetic vacuum fluctuations are fundamental in
many physical processes (spontaneous emission, light scattering, Casimir effect, lasing, etc)~\cite{atomphoton,quantumoptics}
and in a wide variety of applications (quantum information processing, quantum metrology, laser cooling, photonic engineering, etc)~\cite{LiuAdvPhys,OrszagQO,WisemanMilbirnBook}. Particularly in quantum coherence control, the
vacuum fluctuations are usually undesirable~\cite{LiuAdvPhys,OrszagQO,WisemanMilbirnBook,Bergmann_1998,Chen_Raman}
since they cause spontaneous photon emission and in turn irreversible
loss of coherence of the systems under control.
However, there are still some surprising effects.
For instance, it was predicted and observed that
in the stimulated Raman process for spin coherence generation in a $\Lambda$-type
three-level system, the irreversible spontaneous emission (SE) from the optically excited state,
when its spectrum is wide enough to cover both emission pathways and the two pathways couple
to the same photon modes, will generate Raman coherence between different spin states~\cite{Economou_2005,Gurudev:2005PRL}.

In this paper, we predict yet another striking effect of the vacuum fluctuations
and show how it manifests itself in nonlinear optics. The irreversible
SE in a $\Lambda$-type three-level system can cause a spin flip and
hence recover the dephased spin coherence by spin echo~\cite{Hahn}. Such spin-flip control by vacuum fluctuations,
when implemented in the standard spin coherence pump-probe spectroscopy, can realize
spin echo in nonlinear optics, with two orders of optical fields saved as compared with
the conventional methods using stimulated Raman processes.
This effect shows that the vacuum field can indeed replace some orders of
the optical field in nonlinear optical spectroscopy, which is consistent with the
previous results on spin coherence signatures in frequency-domain nonlinear optical spectra~\cite{LiuChi5}.
By studying the nonlinear optical signals of spin coherence in a fluctuating
random field (due to environmental noises)~\cite{Kubo_1969}, we will also show that the SE-assisted
spin flip has the same effect as a usual coherent $\pi$-rotation control in
restoring the spin coherence lost within the memory time of the environmental noises~\cite{Loring:1985}.

The organization of this paper is as follow. After this introductory section, Sec.~\ref{Sec_idea}
describes the basic idea of the paper. Sec.~\ref{model} describes the model for the quantum dot
(QD) system and the master-equation approach to calculating the nonlinear optical susceptibility.
Sec.~\ref{result} presents the results and discussion.
Sec.~\ref{conclusion} concludes this paper. The solution of the master
equation is presented in the Appendix.

\section{Basic idea}
\label{Sec_idea}

\begin{figure}[b]
\begin{center}
\includegraphics[width=\linewidth]{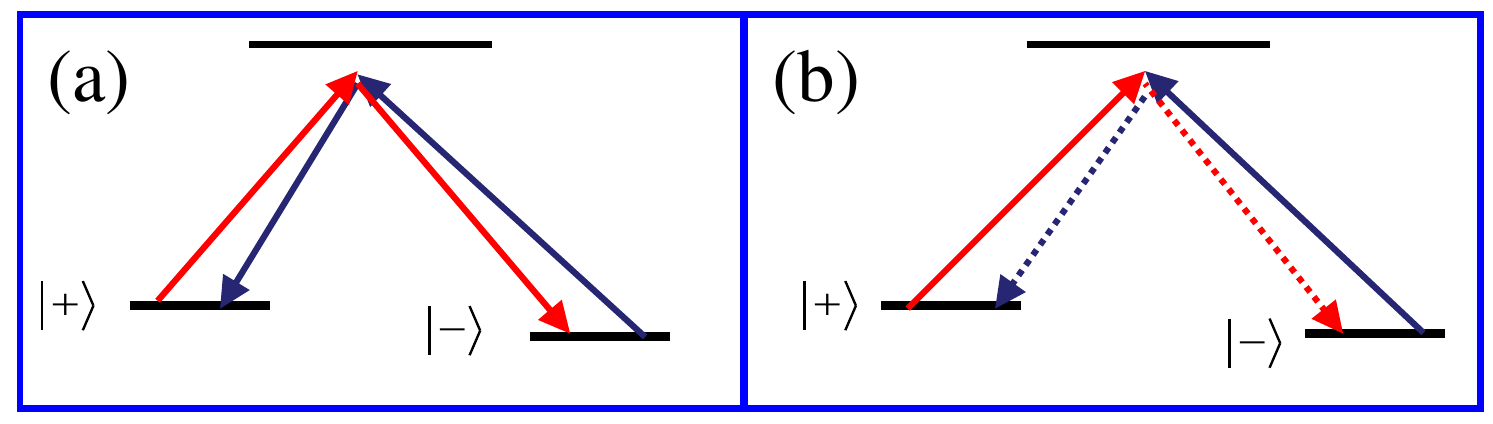}
\end{center}
\caption{(Color online) (a) Stimulated Raman processes for
a spin flip control in a $\Lambda$-type three-level system.
(b) Raman processes for a spin flip control, with emission caused
by vacuum fluctuations (dotted arrows) instead of a laser field as in (a).
}
\label{fig_Raman}
\end{figure}

To illustrate the basic idea of spin flip control by vacuum fluctuations, let us first examine
the stimulated Raman processes. Such processes are the fundamental mechanisms of many physical phenomena
such as electromagnetically induced transparency~\cite{EIT}, stimulated Raman adiabatic
passage~\cite{Bergmann_1998}, and optical control of spins in
semiconductors~\cite{Gupta_2001Science,Wu:2007PRL,Berezovsky:2008,Yamamoto_echo,Greilich2009NPhys,WangHL09SpinFlip,YamamotoQDecho}.
As shown in Fig.~\ref{fig_Raman}~(a), we consider two spin states $|\pm\rangle$ coupled to
the same optically excited state by a short laser pulse. The spin is flipped when $|+\rangle$
and $|-\rangle$ are exchanged, via two state transfer processes in parallel,
namely, the stimulated Raman processes from $|\pm\rangle$ to $|\mp\rangle$ mediated by the optically excited
state. Similar to the photon echo in four-wave mixing~\cite{Lindberg:1992PRA}, the signature
of the spin flip will appear as spin echo in nonlinear optics via a perturbation procedure with four orders
of the optical field involved in the spin flip.

Now if the stimulated photon emission from the optically excited state is replaced by the
SE [see Fig.~\ref{fig_Raman}~(b)], the spin flip
can be realized by Raman processes involving only two orders of the laser field.
Similar to the stimulated Raman processes, it is essential that the SE spectrum
is broader than the spin splitting and the two emission pathways couple to the same photon mode.
Such requirements indicate the fundamental basis of the
predicted effect: The SE (dissipation) is due to the correlation of
the vacuum fluctuations~\cite{Kubo66FD}, so when there are several final states in an SE process,
coherent correlation between different quantum pathways may be generated when the ``which-way'' information
is not fully resolved. Such correlation may lead to coherence generation by SE~\cite{Economou_2005,Gurudev:2005PRL}
and even coherent spin control when there is initial spin coherence.

\section{Model and Theory}
\label{model}

\begin{figure}[b]
\begin{center}
\includegraphics[width=\linewidth]{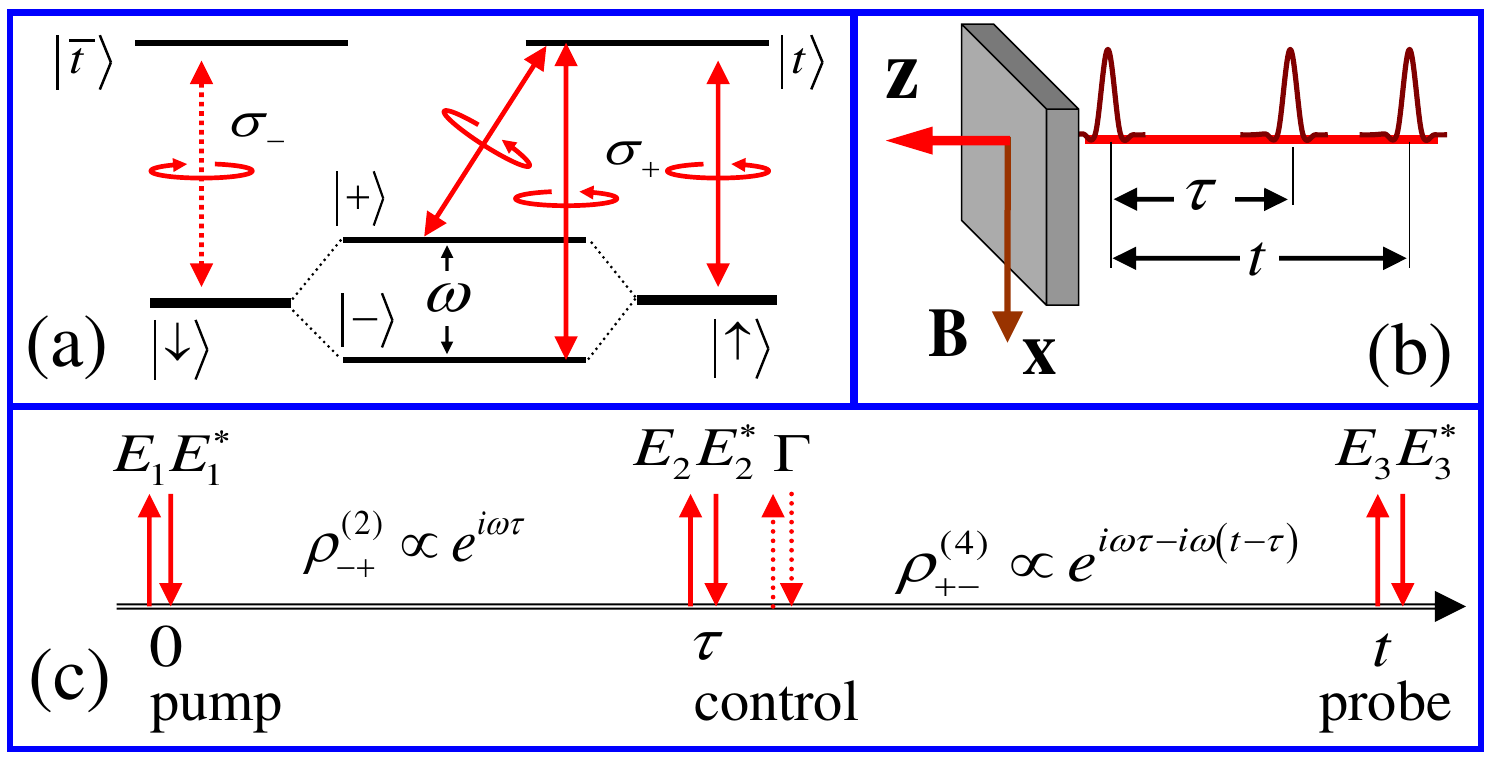}
\end{center}
\caption{(Color online)
(a) Selection rules of optical transitions
in a singly charged fluctuation quantum dot.
(b) Setup for spin echo by nonlinear optics.
(c) Optical couplings (vertical solid arrows) and
    spontaneous emission (vertical dotted arrows) for generation,
    flip control, and detection of spin coherence.
}
\label{fig_model}
\end{figure}

To be specific, we will present the detailed analysis for a model system of electron spins
in QDs, a paradigmatic system in research of quantum optics, quantum computing,
and mesoscopic physics. In a GaAs fluctuation QD doped with a single electron~\cite{Gurudev:2005PRL}, for example,
a normal incident light with circular polarization $\sigma_+$ (or $\sigma_-$) couples
only to the optical transition between the electron spin
state $|\uparrow\rangle$ (or $|\downarrow\rangle$) to the negatively charged
exciton state, i.e., the trion state $|t\rangle$ (or $|\bar{t}\rangle$),
with the spin basis quantized along the growth direction (the $z$-axis) [see Fig.~\ref{fig_model}~(a)].
Under a transverse magnetic field [in the Voigt geometry, see Fig.~\ref{fig_model}~(b)],
the electron spin is split into two states $|\pm\rangle\equiv\left(|\uparrow\rangle\pm|\downarrow\rangle\right)/\sqrt{2}$ quantized along the
external magnetic field direction ($x$-axis), with Zeeman energy $\omega$,
but the trion states remain nearly degenerate due to the
large energy mismatch between the heavy hole and the light hole
and hence can still be quantized along the growth direction~\cite{Gurudev:2005PRL}.
Without loss of generality, we set the pump, control, and probe pulses all $\sigma_+$-polarized.
Then, only the trion state $|t\rangle$ will be excited and thus the system is modeled by a $\Lambda$-type
three-level system consisting of $|\pm\rangle$ and $|t\rangle$.
The all-optical spin echo is based on a standard pump-probe setup [see Fig.~\ref{fig_model}~(b)] and the
basic optical processes are illustrated in Fig.~\ref{fig_model}~(c).

The Hamiltonian of the model system is
\begin{equation}
H=\varepsilon_t|t\rangle\langle t|+\omega S_x+\sum_{j}\left[|t\rangle\langle \uparrow|E_j(t)+\text{h.c.}\right],
\end{equation}
where $\varepsilon_t$ is the energy gap, $\omega S_x$ is the Zeeman coupling with
$S_x\equiv\left( |\uparrow\rangle\langle \downarrow|+|\downarrow\rangle\langle \uparrow|\right)/2$, and $E_j(t)\equiv \chi_j\left(t\right)e^{-i\Omega_jt}$ is the positive-frequency component of
the optical coupling from the $j$th laser pulse with $\chi_j\left(t\right)$ denoting the pulse envelope.
The transition dipole moment is understood to be absorbed into the field quantities.
The first pulse centered at time $0$ prepares the spin coherence,
the second pulse at time $\tau$ realizes the spin control, and the third pulse probes the spin coherence at time $t$.

The nonlinear optical response can be calculated directly by solving the master equation perturbatively
in powers of the optical fields (see Appendix). The master equation reads
\begin{align}
\partial_t\rho=& -i\left[H,\rho \right]
-\left({\Gamma}/{2}\right)\left(\Lambda^{\dag}\Lambda\rho+\rho \Lambda^{\dag}\Lambda -2\Lambda\rho\Lambda ^{\dag} \right)
\nonumber \\
&  -{T_2^{-1}}\left(2S_x\rho S_x-\rho/2\right) ,
\label{Eq:master}
\end{align}
where the irreversible SE of a rate $\Gamma$ is described by the Lindblad form associated with the optical transition $\Lambda\equiv
|\uparrow\rangle\langle t|$, and the pure dephasing of the electron spin is characterized by the $T_2$-term
in the second line.
The extra dephasing due to mechanisms such as phonon scattering and tunneling leakage
is negligible at low temperature and in electrically stable QDs~\cite{Gurudev:2005PRL,Wu:2007PRL,Berezovsky:2008,Greilich2009NPhys,YamamotoQDecho},
and if included, would only quantitatively modify the signal amplitudes without changing the main results of this paper.
The stimulated emission also contributes to the stimulated Raman processes [Fig.~\ref{fig_Raman}~(a)]. This effect,
being proportional to the optical fields, has been automatically included in the coherent driving of the lasers.

The spin splitting $\omega$ consists of the Zeeman energy and the local field fluctuation,
the latter causing the spin dephasing.
At low temperature, the local field fluctuation in QDs
are mainly due to the hyperfine interaction with nuclear spins~\cite{Merkulov:2002PRB}, which includes both
static inhomogeneous broadening and dynamical spectral diffusion.
To simplify the discussions without affecting the essential physics, we will model
the local field fluctuation using a phenomenological random field with certain correlation
functions~\cite{Kubo_1969,Loring:1985}.

\section{Results and Discussion}
\label{result}
\subsection{Spin coherence generation}
The first step is optical pumping of spin coherence.
We assume that the system initially has no spin coherence,
i.e., $\rho=\frac{1}{2}|\uparrow\rangle\langle\uparrow|+\frac{1}{2}|\downarrow\rangle\langle\downarrow|
=\frac{1}{2}|+\rangle\langle+|+\frac{1}{2}|-\rangle\langle-|$.
A short $\sigma_+$-polarized pulse, with a bandwidth greater than the spin splitting, excites population from
the spin state $|\uparrow\rangle$ to the trion state $|t\rangle$, leaving the spins polarized along
the $-z$-direction and initiated to precess about the external field. The SE will bring the trion
population back to the spin state $|\uparrow\rangle$, which tends to cancel the spin coherence generated by the
stimulated Raman processes. As the SE takes a finite time during which
the spins precess, the spin coherence will be only partially canceled and phase delayed~\cite{Economou_2005,Gurudev:2005PRL}.
Starting from the initial population at $|+\rangle$, e.g., the second-order optical processes for the spin coherence generation are described by [see Fig.~\ref{fig_model}~(c)]
\begin{subequations}
\begin{align}
\rho^{(0)}_{++}
& \stackrel{E_1}{\longrightarrow}\rho^{(1)}_{t+}
  \stackrel{E_1^*}{\longrightarrow}\rho^{(2)}_{-+} \ \text{or}\ \rho^{(2)}_{++}, \label{3a}\\
\rho^{(0)}_{++}
& \stackrel{E_1}{\longrightarrow}\rho^{(1)}_{t+}
  \stackrel{E_1^*}{\longrightarrow}\rho^{(2)}_{tt}
\stackrel{\rm SE}{\longrightarrow} \rho^{(2)}_{\uparrow\uparrow},\label{3b}
\end{align}
\end{subequations}
where $\rho^{(n)}_{\alpha\beta}$ is the density matrix element between $|\alpha\rangle$ and $|\beta\rangle$
in the $n$th order of the optical field. The evolution starting from the population $\rho^{(0)}_{--}$ is similar.
The spin coherence after the pump, quantified as the off-diagonal matrix element, is~\cite{Gurudev:2005PRL}
\begin{eqnarray}
\rho^{(2)}_{+-}(t) =\frac{G_1}{2}
\frac{\omega}{\omega+i\Gamma}e^{-i\omega t-t/T_2},
\label{spin_coherence_generation}
\end{eqnarray}
where $T_2$ is the spin decoherence time, and the excitation probability under the resonance condition ($\varepsilon_t=\Omega_1$) is determined by the
pulse area through
\begin{equation}
G_1\equiv \left|\int_{-\infty}^{+\infty} \chi_1\left(t\right) dt \right|^2\propto E_1E_1^*.
\end{equation}
In the presence of inhomogeneous broadening
[a probability distribution of $\omega$ assumed as $e^{-(\omega-\omega_0)^2/(2\sigma^2)}$ around the central frequency $\omega_0$],
the ensemble-averaged spin coherence is
\begin{eqnarray}
\left\langle \rho^{(2)}_{+-}(t) \right\rangle \propto
E_1E_1^*\frac{\omega_0}{\omega_0+i\Gamma}e^{-i\omega_0
t-t/T_2-\sigma^2t^2/2}. \label{spin_coherence_ensemble}
\end{eqnarray}
Here we have used the condition that $\Gamma\gg\sigma$.
As usually $\sigma\gg 1/T_2$, the
spin polarization decay is dominated by the inhomogeneous broadening effect. To resolve the
``true'' spin decoherence, spin echo may be invoked.

\subsection{Spin coherence control}
The key step in the all-optical spin echo is the control of spins.
To illustrate the idea, let us start with the
rotation of spins by the optical AC Stark shift which has been
demonstrated in QDs~\cite{Berezovsky:2008,Greilich2009NPhys,YamamotoQDecho}.
We consider a $\sigma_+$-polarized pulse detuned well below the trion
resonance. The virtual transition between $|t\rangle$ and
the spin state $|\uparrow\rangle$ induces an AC Stark energy shift
of the spin state, which in turn induces a rotation
of the spin about the $z$-axis, with an angle
$
\theta \propto \left|E_2\right|^2+O\left(\left|E_2\right|^4\right).
$
If $\theta=\pi$, the spins are flipped. In reality, it is non-trivial to realize an exact
$\pi$-rotation~\cite{Greilich2009NPhys,YamamotoQDecho}.
The idea of using nonlinear optical response to realize spin echo comes from
the perturbation expansion
\begin{eqnarray}
\exp\left(i\theta S_z\right) = 1 + i\theta S_z + O\left(\theta^2\right).
\label{spin_flip_generator}
\end{eqnarray}
Thus, an infinitesimal rotation contains the rotation generator $S_z$, i.e., the spin operator
along the $z$-axis, which exchanges the states $|+\rangle$ and $|-\rangle$.

From Eq.~(\ref{spin_flip_generator}), it is tempting to conclude that two orders of the control
field can flip the spin coherence. A closer examination,
however, reveals that we need actually four orders of the control field.
To see the problem, let us consider a general spin state
$|\psi\rangle=C_+|+\rangle+C_-|-\rangle$.
A small rotation about the $z$-axis transforms it into
\begin{equation}
e^{iS_z\theta}|\psi\rangle=\left(C_+ +i\frac{\theta}{2}C_-\right)|+\rangle
+\left(C_-+i\frac{\theta}{2}C_+\right)|-\rangle+O\left(\theta^2\right).
\end{equation}
Before the pulse applied at $t=\tau$, the spin coherence is
$\rho_{+-}(\tau-0)=C_+C_-^*\propto \exp\left(-i\omega\tau\right)$.
For spin echo, we wish to pick up the spin-flipped term $\rho_{-+}(\tau+0)$ after the control
pulse. Such a term in the leading order of $\theta$ is $\theta^2 C_+C_-^*/4$.
Thus at least four orders of the control field are needed.
This problem can also be understood from the picture of stimulated Raman processes
shown in Fig.~\ref{fig_Raman}~(a) or from the excitation pathways of the control process
(see formula below).
Starting from the spin coherence generated by the pump pulse,
$\rho^{(2)}_{+-}$, the excitation by two orders of the control pulse
follows the pathways
\begin{subequations}
\begin{align}
\rho^{(2)}_{+-}&
\stackrel{E_2}{\longrightarrow}\rho^{(3)}_{t-}
\stackrel{E_2^*}{\longrightarrow}
\rho^{(4)}_{--},\ \rho^{(4)}_{+-},\ {\rm or}\ \rho^{(4)}_{tt},\label{8a}
\\
\rho^{(2)}_{+-}&
\stackrel{E_2^*}{\longrightarrow}\rho^{(3)}_{+t}
\stackrel{E_2}{\longrightarrow}
\rho^{(4)}_{++},\ \rho^{(4)}_{+-},\  {\rm or}\ \rho^{(4)}_{tt},\label{8b}
\end{align}
\label{pathways}
\end{subequations}
none of which results in a spin-flipped term $\rho_{-+}^{(4)}$.  Note that the excitation pathways
are independent of the detuning of light, and thus the problem discussed above
is not limited to the spin rotation by the AC Stark effect of virtual excitation but
applies also to real excitation.

We note that in Eq.~(\ref{pathways}) the trion population is also obtained if the excitation is
in resonance with the trion. As discussed earlier for the optical pump of spin coherence, the SE
will bring the trion population to the spin population $\rho^{(4)}_{\uparrow\uparrow}$.
Thus with the SE included, the spin-flipped coherence is obtained through
the quantum pathway [see Fig.~\ref{fig_model}~(c)]
\begin{equation}
\rho^{(2)}_{+-}
\stackrel{E_2 E_2^*}{\longrightarrow}
\rho^{(4)}_{tt}\stackrel{\rm SE}{\longrightarrow}
\rho^{(4)}_{\uparrow\uparrow}=\frac{1}{2}\left(\rho^{(4)}_{++}+\rho^{(4)}_{-+}+\rho^{(4)}_{+-}+\rho^{(4)}_{--}\right).
\label{pathways_SE}
\end{equation}
Indeed, one can regard the SE as the contribution of two orders of the vacuum field to the
nonlinear optical response, which is consistent with the observation that at least
four orders of control field are needed to flip the spin coherence.
Similar to the stimulated Raman processes, we also need the bandwidth of the SE to be comparable
to or greater than the spin splitting (i.e., $\Gamma\gtrsim \omega$) .

Considering the spin precession during the SE, the spin coherence generated
by the SE is~\cite{Gurudev:2005PRL}
\begin{eqnarray}
\rho^{{\rm SE}(4)}_{-+}(t)=\frac{1}{2}\rho^{(4)}_{tt}\frac{i\Gamma}{\omega-i\Gamma}e^{i\omega(t-\tau)-(t-\tau)/T_2},
\label{coherence_SE}
\end{eqnarray}
where the 4th order trion population is
\begin{eqnarray}
\rho^{(4)}_{tt}=G_2\rho^{(2)}_{\uparrow\uparrow}=\frac{G_2}{2}
\left(\rho^{(2)}_{++}+\rho^{(2)}_{--}+\rho^{(2)}_{+-}+\rho^{(2)}_{-+}\right),
\label{chiSE}
\end{eqnarray}
with the excitation probability under the resonance condition ($\varepsilon_t=\Omega_2$)
\begin{equation}
G_2\equiv \left|\int_{-\infty}^{+\infty} \chi_2\left(t\right) dt \right|^2\propto E_2E_2^*.
\end{equation}
Thus we obtain the spin-flipped coherence term
\begin{eqnarray}
\bar{\rho}^{(4)}_{-+}(t)=\frac{G_2}{4}
\frac{i\Gamma}{\omega-i\Gamma}
e^{i\omega\left(t-\tau\right)-(t-\tau)/T_2}\rho^{(2)}_{+-}(\tau).
\label{flip_coherence}
\end{eqnarray}
With the spin coherence generated by the pump pulse given in Eq.~(\ref{spin_coherence_generation}),
the ensemble average of the spin-flipped term is
\begin{eqnarray}
\left\langle\bar{\rho}^{(4)}_{-+}(t)\right\rangle \propto
\left|E_1\right|^2\left|E_2\right|^2 \frac{i\omega_0\Gamma}{\omega_0^2+\Gamma^2}
\left\langle e^{i\omega(t-2\tau)-t/T_2}\right\rangle.
\label{spin_echo}
\end{eqnarray}
The spin echo is seen by noticing that the phase factor $e^{-i\omega\tau}$ accumulated in
$\rho^{(2)}_{+-}(\tau)$ is canceled in $\bar{\rho}^{(4)}_{-+}(t)$ at $t=2\tau$.

\begin{figure}[t]
\begin{center}
\includegraphics[width=0.37\linewidth]{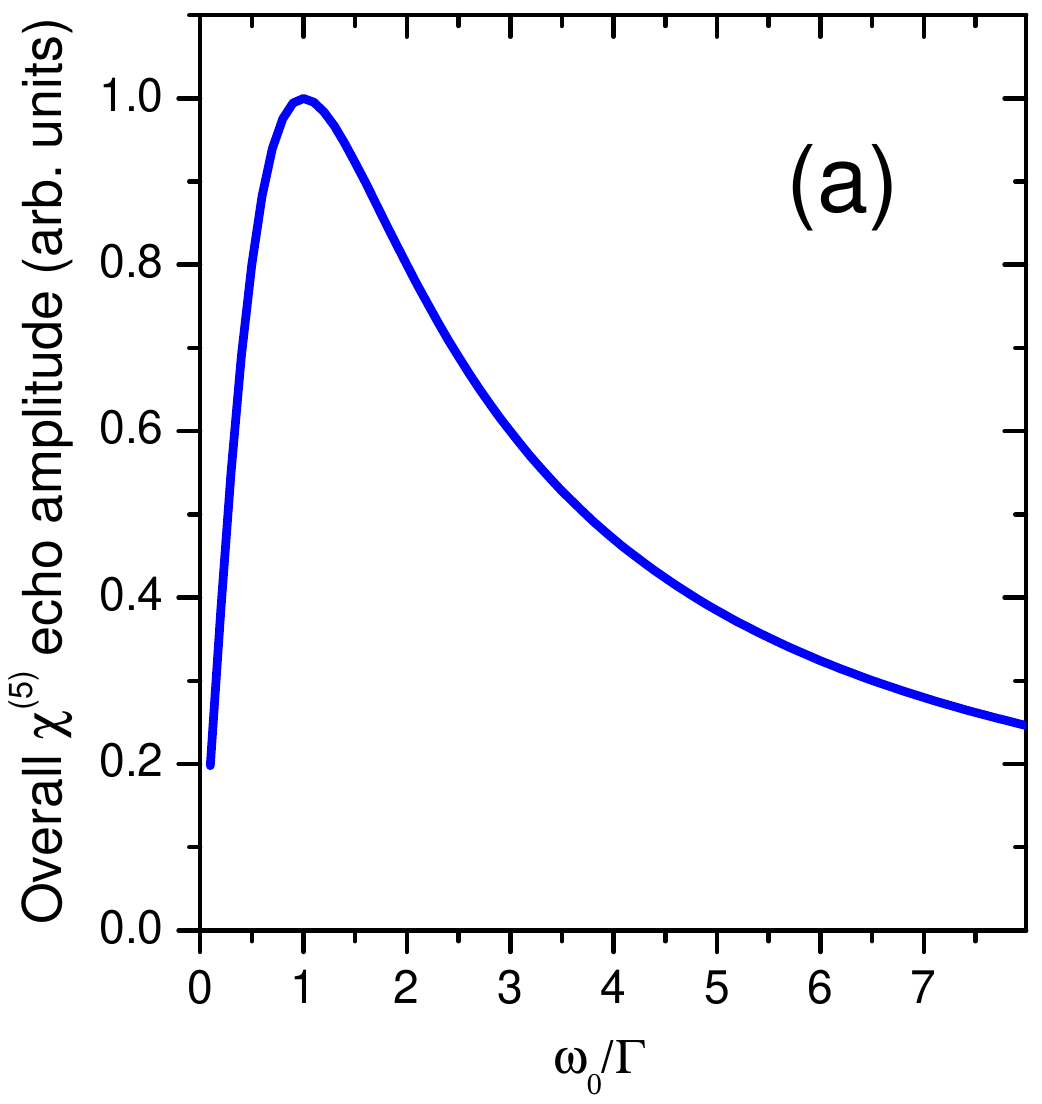}
\hfill
\includegraphics[width=0.59\linewidth]{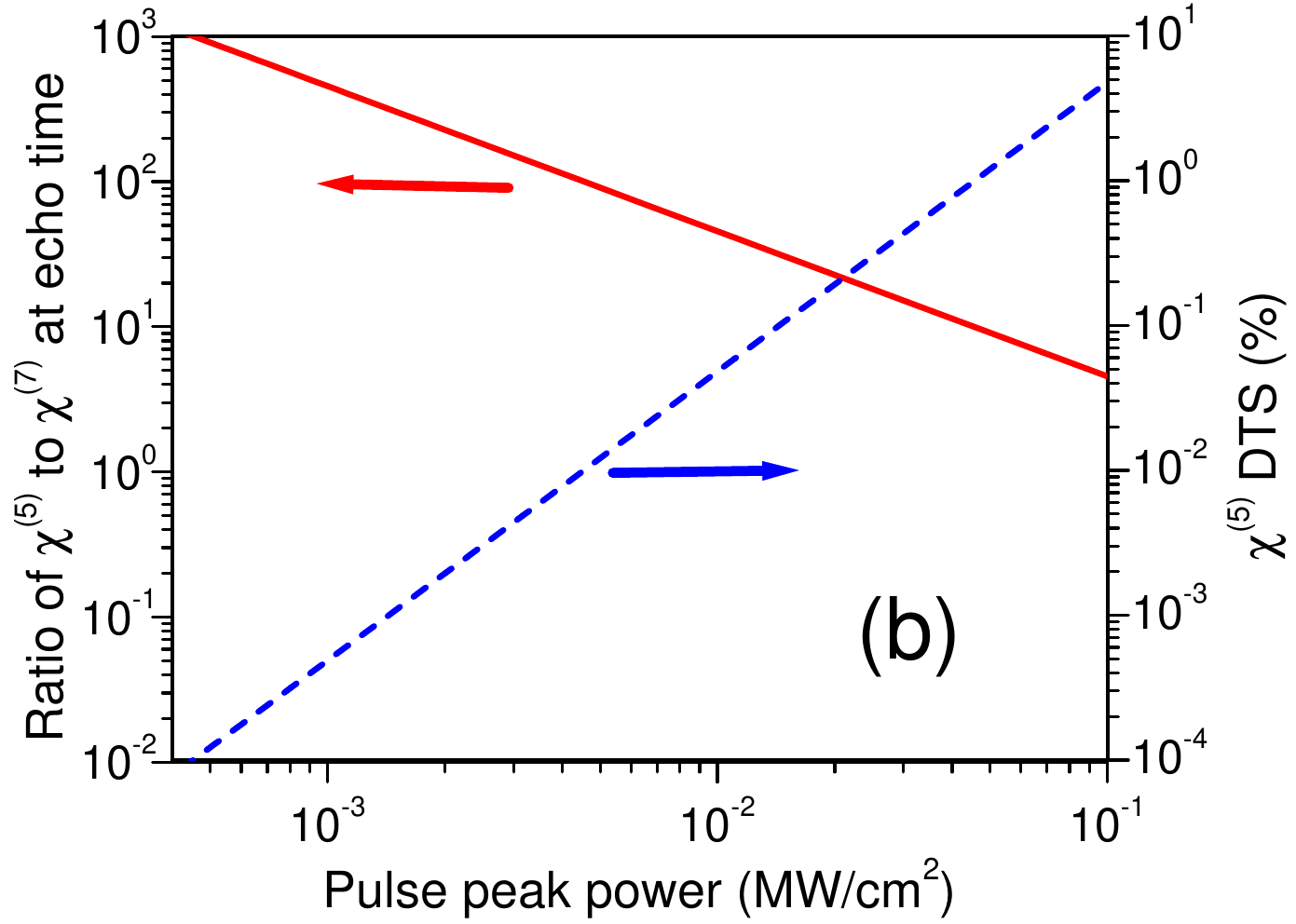}
\end{center}
\caption{(Color online) (a) Overall spin echo amplitude as a function of the ratio of the Zeeman splitting to
the spontaneous emission rate ($\omega_0/\Gamma$). (b) The solid line is the estimated amplitude of the echo induced by the spontaneous emission
relative to that induced by direct laser pulse control [$R_{5/7}$ in Eq.~(\ref{R57})],
and the dashed line is the estimated $\chi^{(5)}$ differential transmission (in percentage of the absorption
without the pump and control pulses)
at the echo time [Eq.~(\ref{DTS5})], both plotted as a function
of the laser pulse intensity. For the estimation,
the pump and the control pulses have the same amplitude ($G_1=G_2$) and the same duration (1~picosecond),
the dipole moment of the exciton is 75~debye (15~e\AA), the dielectric constant
of the material is 10, and $\omega_0=\Gamma$ is assumed.
}
\label{fig_amp}
\end{figure}

The relative magnitude of the echo signal depends on the ratio of the spin splitting $\omega_0$ to the SE rate $\Gamma$.
In generation of the spin coherence by the first pump pulse, faster SE would lead to weaker spin coherence,
but in the spin flip control by the second pulse, faster SE would induce more spin coherence flipped.
Competition between the two effects makes the overall amplitude of the echo signal
peak at $\omega_0/\Gamma=1$ and decreasing to zero when $\omega_0/\Gamma\rightarrow 0$ or $\infty$ [see Fig.~\ref{fig_amp}~(a)].

For comparison, we also consider the direct spin coherence control by the laser pulse without involving the SE.
As discussed in the beginning of this subsection, at least four orders of the optical field
are needed to realize the spin flip. The spin coherence directly flipped by the laser pulse in the leading order
is
\begin{align}
\bar{\rho}^{(6)}_{-+}(t)=\frac{g_2}{4}\rho^{(2)}_{+-}(\tau)e^{i\omega\left(t-\tau\right)-(t-\tau)/T_2},
\label{chi6}
\end{align}
with
\begin{equation}
g_2\equiv \left|\int_{-\infty}^{+\infty} \chi_2^*\left(t'\right) d t'\int^{t'}_{-\infty}\chi_2\left(t\right)dt\right|^2=\frac{1}{4}G_2^2,
\end{equation}
where in the equation above we have used the resonance condition ($\varepsilon_t=\Omega_2$) and assumed
that the pulse envelope $\chi_2(t)$ is a real function.
The relative strength of the echo signal induced by the SE [in Eq.~(\ref{chiSE})]
as compared with that induced by the laser pulse [in Eq.~(\ref{chi6})] is
\begin{equation}
R_{5/7}=\frac{4}{G_2\sqrt{1+\omega_0^2/\Gamma^2}}.
\label{R57}
\end{equation}
This ratio is plotted in Fig.~\ref{fig_amp}~(b) as a function of the laser pulse intensity.
For $\Gamma\sim\omega_0$, the SE-induced echo signal will dominate the higher-order signal induced by
the laser pulse, since $G_2\ll 1$ is satisfied in the perturbative response regime. For example,
for GaAs fluctuation QDs with exciton dipole moments of 75~debye (i.e., 15~e\AA)~\cite{Rabi},
to achieve $G_2\sim 1$, a laser pulse of one picosecond duration needs to have a peak energy flux to be
as high as $0.5$~MW/cm$^2$, or there should be as many as about $2\times 10^3$ photons per pulse per QD
for a QD density of $10^9$~cm$^{-2}$.
Such surprisingly high nonlinearity induced by the electromagnetic vacuum field
can be understood by virtue of the fact that the laser approaches to the QD in only one mode,
while all modes of the vacuum field participate in the SE process.

\subsection{Differential transmission signal}

The differential transmission of a $\sigma_+$-polarized pulse probes the population change of the spin
state $|\uparrow\rangle$ due to the pump and the control pulses.
With two orders of the pump field and two orders of the control field carried by the spin coherence,
the optical polarization induced by the probe pulse is a 5th-order optical response
\begin{align}
\rho^{(5)}_{t\uparrow}(t)=-i\int_{-\infty}^{t}\rho^{(4)}_{\uparrow\uparrow}(t')\chi_3\left(t'\right)dt'.
\end{align}
In heterodyne detection~\cite{heterodyne}, the differential transmission (DTS) of the probe pulse arriving at time $t$ is
\begin{align}
\Delta T^{(5)} \left(t\right) = G_3^{-1}  \text{Im}\left[ \int_{-\infty}^{+\infty} \chi_3^*\left(t'\right)\rho^{(5)}_{t\uparrow}(t') dt'\right],
\end{align}
where the absorption of the probe pulse in absence of the pump and control pulses is
\begin{equation}
G_3\equiv\frac{1}{2}\left|\int_{-\infty}^{+\infty} \chi_3(t)dt\right|^2.
\end{equation}
With the spin coherence given by
\begin{equation}
\rho^{(4)}_{\uparrow\uparrow}(t)
=\left(\rho^{(4)}_{++}+\rho^{(4)}_{--}+2\text{Re}\rho^{(4)}_{+-}\right)/2,
\end{equation}
the DTS is
\begin{align}
\Delta T^{(5)}(t)= C+I_{\text{echo}}\text{Re} \left\langle e^{i\omega(t-2\tau)-t/T_2}\right\rangle,
\label{DTS}
\end{align}
where $C$ consists of all the background terms and the oscillation terms without spin flip, and the
spin echo signal strength is
\begin{equation}
I_{\text{echo}}=\frac{G_1G_2}{4}\frac{ \omega_0 \Gamma}{\Gamma^2+\omega_0^2}.
\label{DTS5}
\end{equation}
Fig.~\ref{fig_amp}~(b) shows the dependence of the signal strength on the laser power under
realistic experimental conditions. For an ensemble with inhomogeneous broadening $\sigma$,
the signal at a long time ($t\gg\sigma^{-1}$) will present oscillation only near the
echo time $t=2\tau$. The decay of the echo signal as a function of $\tau$ reveals the
``true'' decoherence excluding the inhomogeneous broadening effect.

\begin{figure}[b]
\begin{center}
\includegraphics[width=\linewidth]{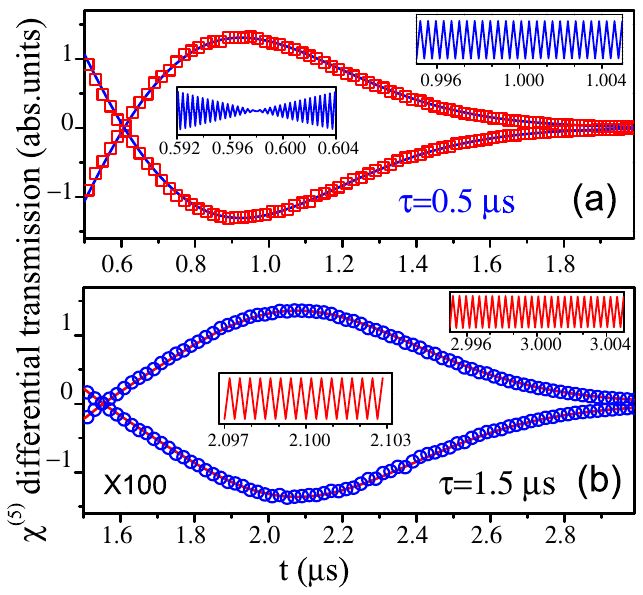}
\end{center}
\caption{(Color online) $\chi^{(5)}$ DTS of singly-charged QDs,
calculated analytically using Eq.~(\ref{Gaussian}) (lines) or numerically (symbols).
(a) and (b) are the envelopes of real-time signals with the control pulse applied
at $\tau=0.5$~$\mu$s or $\tau=1.5$~$\mu$s [signal in (b) amplified by 100], respectively,
with insets showing the oscillations in a few small time-windows.
Corresponding to realistic conditions, the parameters are chosen
as $\tau_{\text c}=1$~$\mu$s, $\omega_0 = 10$~$\mu$eV,
$T_2 \equiv\left\langle X^2(0)\right\rangle^{-1}\tau_{\text c}^{-1} =0.1$~$\mu$s,
and $\Gamma=10$~$\mu$eV. The inhomogeneous broadening
is artificially set to zero.
}
\label{fig_DTS}
\end{figure}

\begin{figure}[t]
\begin{center}
\includegraphics[width=0.9\linewidth]{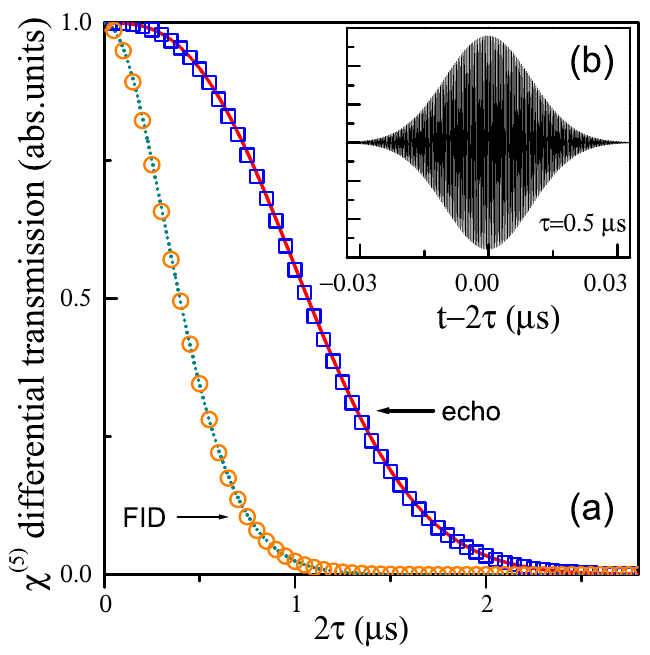}
\end{center}
\caption{(Color online) $\chi^{(5)}$ DTS of singly-charged QDs,
calculated analytically using Eq.~(\ref{Gaussian}) (lines) or numerically (symbols).
(a) is the spin echo signal at $t=2\tau$, with
the free-induction decay (FID) without the inhomogeneous broadening
at $t=2\tau$ plot for comparison (dotted line with circle symbols).
(b) shows the real-time dependence of the signal near the echo time for $\tau=0.5$~$\mu$s.
The parameters are the same as in Fig.~\ref{fig_DTS}, but the inhomogeneous broadening
$\sigma=0.1$~ns$^{-1}$.
}
\label{fig_echo}
\end{figure}

\subsection{Numerical results}
To check whether the spin-flip control by SE can suppress the decoherence in a ``slow'' bath the same way as
a coherent $\pi$-rotation in spin echo~\cite{Loring:1985}, we simulate the decoherence by a spectral diffusion
model in which the local magnetic field $\omega(t)=\omega+X(t)$ contains a dynamically
fluctuating part $X(t)$~\cite{Kubo_1969}.
The accumulated random phase $\phi({t_2},{t_1})\equiv \int^{t_2}_{t_1}X(t)dt $ causes the spin decoherence.
For a Gaussian fluctuation to which Wick's theorem applies,
the spin-flipped coherence term in Eq.~(\ref{spin_echo}) becomes~\cite{Kubo_1969,Loring:1985}
\begin{eqnarray}
\left\langle\bar{\rho}^{(4)}_{-+}(t)\right\rangle \propto
e^{i\omega_0(t-2\tau)-\sigma^2(t-2\tau)^2/2-
\left\langle\left[\phi\left(t,\tau\right)-\phi\left(\tau,0\right)\right]^2\right\rangle/2}.
\label{Gaussian}
\end{eqnarray}
To be specific, we employ a noise correlation of the form~\cite{Loring:1985}
$\left\langle X(t_1)X(t_2)\right\rangle=\left\langle X^2(0)\right\rangle
\exp\left(-\left|t_1-t_2\right|/\tau_{\text c}\right)$.
The spin echo not only eliminates the inhomogeneous broadening effect but also
partially suppress the decoherence resulting from the dynamical fluctuation if the pulse
delay time $\tau$ is comparable to or shorter than the noise correlation time $\tau_{\text c}$~\cite{Loring:1985}.

The partial recovery from the spin decoherence is seen in Fig.~\ref{fig_DTS},
which is obtained by numerical solution of Eq.~(\ref{Eq:master}).
To show the effect of dynamical fluctuation, the inhomogeneous broadening $\sigma$ is artificially
set to zero in Fig.~\ref{fig_DTS}.
When the pulse delay time is shorter than the noise correlation time
[Fig.~\ref{fig_DTS} (a)],
the coherence is recovered near $t=2\tau$, the same as in spin echo for
inhomogeneous broadening which can actually be understood as
spectral diffusion with infinite correlation time~\cite{Loring:1985}.
For longer pulse delay times, the recovery is less perfect and
the peak time approaches to $t=\tau+\tau_{\text c}\ln 2$ (as derived in Ref.~\onlinecite{Loring:1985}),
as evidenced in Fig.~\ref{fig_DTS}~(b).

When the inhomogeneous broadening is included, the signal for $\tau \gg 1/\sigma$
is visible only near the echo time $2\tau$, as shown in Fig.~\ref{fig_echo}~(b).
Fig.~\ref{fig_echo}~(a) plots the echo signal
as a function of the pulse delay time. When $\tau \lesssim\tau_{\text c}$,
the spin coherence lost by the dynamical fluctuation is partially
recovered, and the echo signal decays slower than the free-induction
decay signal [$\propto \text{Re}\left\langle e^{i\phi(2\tau,0)}\right\rangle$]
without the inhomogeneous broadening ($\sigma=0$).

\section{Conclusion}
\label{conclusion}

We have discovered a striking effect of correlation between different quantum pathways
of spontaneous emission in a $\Lambda$-type three-level system, namely, the coherent spin control
by SE and its role in all-optical spin echo. It is shown that two orders of optical field can be
replaced by the vacuum field in the nonlinear optical spectroscopy of spin coherence.
The effect, demonstrated in this paper for spins in quantum dots, should exist in general
two-level systems with splitting comparable to the rate of emission from an excited state,
and may be induced by other dissipation processes such as phonon emission.
It is conceivable that in more general multi-level systems, (higher order) correlations
between multiple decay pathways could lead to a wealth of new physics.

\acknowledgments
This work was supported by the Hong Kong RGC CUHK402207, CUHK Focused
Investments Scheme, National Natural Science
Foundation of China Nos. 11028510, 10774086, 10574076, and 11074143, the Basic Research Program
of China Grant Nos. 2006CB921500 and 2011CB921901, and the US NSF 0804114.
We are grateful to Z. J. Xu for help on the numerical simulation.

\appendix
\section{Perturbative solution of master equation}
The master equation of the density matrix elements in the basis of $\left\{|t\rangle,|+\rangle, |-\rangle\right\}$ in Eq.~(\ref{Eq:master})
is expanded in powers of the optical fields as
\begin{subequations}
\begin{align}
\partial_t\rho^{(2n+1)}_{t,\pm} =& -i\left(\varepsilon_t\mp\omega/2-i{\Gamma}/{2}\right)
\rho^{(2n+1)}_{t,\pm}
\nonumber \\ & -iE(t)\rho^{(2n)}_{t,t}+iE(t)\rho^{(2n)}_{\mp,\pm}+iE(t)\rho^{(2n)}_{\pm,\pm},\\
\partial_t\rho^{(2n)}_{t,t} =& -\Gamma\rho_{t,t}^{(2n)} + 2\text{Im}\left[E^*(t)\rho^{(2n-1)}_{t,+}+E^*(t)\rho^{(2n-1)}_{t,-}\right],\\
\partial_t\rho^{(2n)}_{\pm,\pm} =& \left({\Gamma}/{2}\right)\rho_{t,t}^{(2n)}-2\text{Im}\left[E^*(t)\rho^{(2n-1)}_{t,\pm}\right],\\
\partial_t\rho^{(2n)}_{\pm,\mp} =& \left({\Gamma}/{2}\right)\rho^{(2n)}_{t,t}  \mp i\omega\rho_{\pm,\mp}^{(2n)}-{T_2^{-1}}\rho_{\pm,\mp}^{(2n)}
\nonumber \\ & + iE^*(t)\rho^{(2n-1)}_{t,\mp} -iE(t)\rho^{(2n-1)}_{\pm,t},
\end{align}
\end{subequations}
where $E(t)\equiv \sum_jE_j(t)$.
In the rotating wave reference frame, the energy gap $\varepsilon_t$ is set to be zero and the optical frequency $\Omega_j$ are measured from the gap.
We assume that the initial density matrix in the equilibrium state is
\begin{align}
\rho^{(0)}=\frac{1}{2}|\uparrow\rangle\langle\uparrow|+\frac{1}{2}|\downarrow\rangle\langle\downarrow|=
\frac{1}{2}|+\rangle\langle+|+\frac{1}{2}|-\rangle\langle-|,
\end{align}
i.e., there is no spin coherence.
The master equation can be solved perturbatively in the order of optical fields,
\begin{align}
\rho^{(0)}
& \stackrel{E_1}{\longrightarrow}\rho^{(1)}\stackrel{E^*_1}{\longrightarrow}\rho^{(2)}
\stackrel{E_2}{\longrightarrow}\rho^{(3)}\stackrel{E^*_2}{\longrightarrow}\rho^{(4)}\stackrel{E_3}{\longrightarrow}\rho^{(5)}.
\end{align}
The derivation of the density matrix elements up to the fifth order is lengthy but straightforward.

We consider an ultra-short optical pulse exciting electrons from $|\uparrow\rangle$ to the trion state $|t\rangle$. Such an
excitation can be taken as instantaneous. Right after the pulse excitation, the
second-order density matrix can be formulated in the Lindblad form as
\begin{align}
\rho^{(0)} & \stackrel{\text{excitation}}{\longrightarrow}\rho^{(2)} = - \frac{G_1}{2}\left[\Lambda \Lambda^{\dag}\rho^{(0)}+
\rho^{(0)} \Lambda^{\dag}\Lambda-2\Lambda\rho^{(0)}\Lambda^{\dag}\right]
\nonumber\\
=&\frac{G_1}{2}|t\rangle\langle t|-\frac{G_1}{2}|\uparrow\rangle\langle\uparrow|\nonumber\\
=&\frac{G_1}{2}|t\rangle\langle t|-\frac{G_1}{4}\left(|+\rangle\langle+|+|-\rangle\langle-|+|+\rangle\langle-|+|-\rangle\langle+|\right),
\label{excitation}
\end{align}
where the excited trion population $G_1/2\propto E_1E_1^*$. After the excitation, a portion of population ($G_1/2$)
at the $|\uparrow\rangle$ state is moved to the trion state $|t\rangle$,
and the spin population is unbalanced in the $z$-axis. Thus off-diagonal coherence in the $x$-basis is generated.

Now let us consider the SE. The Lindblad form for the SE is given in Eq.~(\ref{Eq:master}).
If the SE is much faster than the spin precession,
the trion would return to the spin state $|\uparrow\rangle$ immediately after the excitation. The
induced second-order density matrix (obtained by direct integration of the master equation) is
\begin{align}
\rho^{(2)}& \stackrel{\text{emission}}{\longrightarrow}\rho^{(2)\prime} = \rho^{(2)}
-\frac{1}{2}\left[\Lambda^{\dag}\Lambda\rho^{(2)}+\rho^{(2)} \Lambda^{\dag}\Lambda-2\Lambda\rho^{(2)} \Lambda^{\dag}\right],\nonumber\\
=&0.
\end{align}
The spin coherence is canceled. In this extreme case, the optical process in Eq.~(\ref{3a}) can not generate spin coherence.

In reality, the spontaneous emission has a finite life time $1/\Gamma$, so the spin population returning to
the $|\uparrow\rangle$ state at different times would precess with different phaseshifts and
 the summation would not cancel the spin coherence generated by the optical excitation.
 The spontaneous emission during a finite time can be described by the quantum jump theory as~\cite{Economou_2005}
\begin{align}
\rho^{(2)\prime}(t) &= \mathcal{U}(t)
\left[\rho^{(2)}\right]- \int_0^t\mathcal{U}(t-t')\mathcal{L}\mathcal{U}(t')\left[\rho^{(2)}\right]e^{-\Gamma t'}\Gamma dt',
\end{align}
where $\mathcal{L}\left[\rho\right]\equiv\Lambda^{\dag}\Lambda\rho^{(2)}+\rho^{(2)} \Lambda^{\dag}\Lambda-2\Lambda\rho^{(2)} \Lambda^{\dag}$
is the Lindblad form for the emission and $\mathcal{U}(t)\left[\rho\right]\equiv\exp\left(-i\omega tS_x\right)\rho\exp\left(i\omega tS_x\right)$ is the spin precession process. The net off-diagonal spin coherence in the $x$-basis would be reduced and phase shifted to be
\begin{align}
\rho_{+-}^{(2)}(t) &= \frac{G_1}{2}\left(1-\frac{\Gamma}{\Gamma-i\omega}\right)e^{-i\omega t-t/T_2}
\nonumber\\
&\propto E_1E_1^*\frac{\omega}{\omega+i\Gamma}e^{-i\omega t-t/T_2},\label{SE}
\end{align}
where we have included the spin decoherence time $T_2$ which comes from the dynamical fluctuation of $\omega$. Thus
spin coherence is generated through SE in the optical process illustrated in Eq.~(\ref{3b}).

To realize the spin echo, we wish to transform the off-diagonal spin coherence $\rho^{(2)}_{+-}$ to be $\rho^{(4)}_{-+}$ after
a pulse of quantum control. Let us first consider the excitation.
The leading order effect of the excitation on the spin coherence is a small removal of the spin
population from $|\uparrow\rangle$, $G_2$, which is proportional to the light intensity. By the Lindblad form for the
excitation as in Eq.~(\ref{excitation}), the change of the density matrix is
\begin{align}
\rho^{(4)} &= G_2\left(\rho^{(2)}_{\uparrow\uparrow}|t\rangle\langle t|
-\rho^{(2)}_{\uparrow\uparrow}|\uparrow\rangle\langle\uparrow|
-\frac{1}{2}\rho^{(2)}_{\uparrow\downarrow}|\uparrow\rangle\langle\downarrow|
-\frac{1}{2}\rho^{(2)}_{\downarrow\uparrow}|\downarrow\rangle\langle\uparrow|
\right),
\end{align}
the off-diagonal term in the $x$-basis is
\begin{align}
\rho^{(4)} _{+,-} &= -\frac{G_2}{2}\left(\rho^{(2)}_{\uparrow\uparrow}-\frac{1}{2}\rho^{(2)}_{\uparrow\downarrow}
+\frac{1}{2}\rho^{(2)}_{\downarrow\uparrow}\right)
\nonumber\\
&=-\frac{G_2}{4}\left(\rho^{(2)}_{++}+\rho^{(2)}_{--}+2\rho^{(2)}_{+-}\right)
\end{align}
There is no spin-index flip for the off-diagonal term ($\rho^{(2)}_{-+}\rightarrow\rho^{(4)}_{+-}$). Thus no spin-echo could be
realized through optical processes in Eq.~(\ref{8a}) or (\ref{8b}).

The SE brings the trion population to the spin state $|\uparrow\rangle$, which generate the
off-diagonal spin coherence in the $x$-basis. For simplicity, let us first consider instantaneous SE, the resultant state is
\begin{align}
\bar{\rho}^{(4)} &= G_2\rho^{(2)}_{\uparrow\uparrow}|\uparrow\rangle\langle\uparrow|,
\end{align}
which in the $x$-basis is
\begin{align}
\bar{\rho}^{(4)}_{+-}= \frac{G_2}{2}\rho^{(2)}_{\uparrow\uparrow}=
\frac{G_2}{4}\left(\rho^{(2)}_{++}+\rho^{(2)}_{--}+\rho^{(2)}_{+-}+\rho^{(2)}_{-+}\right).
\end{align}
Thus we get a spin-index flipped term, which contributes to the spin echo signal.
For a finite SE time, the spin coherence generated by the SE can be derived with
the quantum jump theory as given in Eq.~(\ref{SE}). The contribution to the spin
echo signal is
\begin{align}
\bar{\rho}^{(4)}_{+-}(t) = & \frac{G_2}{4}\frac{\Gamma}{\Gamma-i\omega}e^{-i\omega(t-\tau)-(t-\tau)/T_2}\rho^{(2)}_{-+}.
\end{align}
Thus the optical processes in Eq.~(\ref{pathways_SE}) realizes the spin echo.


\end{document}